\documentclass[11pt]{article}\pdfoutput=1
\usepackage{cite}
\usepackage{amsmath,amsfonts,amssymb}%,calrsfs}
\usepackage[small,bf,hang]{caption}
\usepackage{latexsym,epsfig}
\usepackage{arydshln}

\usepackage{chngcntr}

\def\hybrid{
        \topmargin -20pt
        \oddsidemargin 0pt
        \headheight 0pt \headsep 0pt
        \textwidth 6.25in % A4 paper
        \textheight 9.5in % A4 paper
        \marginparwidth .875in
        \parskip 5pt plus 1pt \jot = 1.5ex}

\hybrid

 \csname
@addtoreset\endcsname{equation}{section}

\def\moth{\mathsurround=0pt}
\newdimen\zo \zo=0pt

\def\tick{\leaders\hrule height 0.5ex depth 0pt \hskip 0.5pt}
\def\upboxfill{$\moth \setbox\zo\hbox{\tick}%
  \hskip 3pt\hbox to 0pt{$\tick$\hss}\hrulefill \hbox to 7.5pt{$\tick$\hss}$}

\def\dtick{\leaders\hrule height .34pt depth 0.5ex \hskip 0.5pt}
\def\downboxfill{$\moth \setbox\zo\hbox{\dtick}%
  \hskip 2pt\hbox to 0pt{$\dtick$\hss}\hrulefill \hbox to 2pt{$\dtick$\hss}$}

\def\S{{\cal S}}

\def\bec{\begin{center}}
\def\ec{\end{center}}

\def\tr{{\rm tr}}

 \def\det{{\rm det\,}}
\def\be{\begin{equation}}
\def\ee{\end{equation}}
\def\bea{\begin{eqnarray}}
\def\eea{\end{eqnarray}}
\def\ba{\begin{array}}
\def\ea{\end{array}}

\begin{document}

\begin{titlepage}
\rightline{}
%\rightline\today
\rightline{Submitted March 31, 2019}
\begin{center}
\vskip 2cm
{\Large \bf{Non-perturbative de Sitter vacua via $\alpha'$ corrections}
}\\
\vskip 2.2cm

{\large\bf {Olaf Hohm\footnote{Corresponding author}${}^{,\,\bot}$ and Barton Zwiebach${}^{\top}$}}
\vskip 1.6cm

{\it ${}^{\bot}$ Institute for Physics, Humboldt University Berlin,\\
 Zum Gro\ss en Windkanal 6, D-12489 Berlin, Germany}\\[0.5ex]
ohohm@physik.hu-berlin.de\\ 
\vskip .1cm

\vskip .2cm

{\em $^{\top}$ \hskip -.1truecm Center for Theoretical Physics, \\
Massachusetts Institute of Technology\\
Cambridge, MA 02139, USA
}\\[0.5ex]
zwiebach@mit.edu

\end{center}

\bigskip\bigskip
\begin{center} 
\textbf{Abstract}

\end{center} 
\begin{quote}

The higher-derivative $\alpha'$ corrections 
consistent with $O(d,d)$ duality invariance can be completely
classified for  cosmological, purely time-dependent backgrounds. 
This result is used to show that there are duality invariant theories featuring 
string-frame 
de Sitter vacua as solutions that are non-perturbative in $\alpha'$, 
thus suggesting that  classical   
string theory may realize de Sitter solutions in an unexpected fashion.

\end{quote}

\bigskip\bigskip\bigskip\bigskip

\noindent Second Prize Essay for the Gravity Research Foundation 2019 Awards for Essays on Gravitation
 
\vfill

\end{titlepage}

%\setcounter{tocdepth}{1}
%\tableofcontents

%\newpage 

\counterwithout{equation}{section}

\setcounter{equation}{0}

Since the discovery in the late 1990s that the Universe 
is presently undergoing 
accelerated expansion it has been a challenge for fundamental theories such as string theory 
to provide a natural explanation for this phenomenon. The arguably most convincing explanation 
is a positive cosmological constant leading to approximate de Sitter spacetimes. 
Accordingly, the challenge has been to find de Sitter solutions in string theory. 
But despite more than two decades of efforts not a single uncontested construction of 
de Sitter vacua in string theory has emerged. 
It  has become a question of principle  whether 
string theory permits 
 de Sitter solutions. 
Recently, this state of affairs has led some researchers 
to conjecture that there are no de Sitter vacua in string theory and that, if string theory is to remain tenable, 
one has to search for other explanations of the accelerated expansion of the Universe \cite{Obied:2018sgi,Agrawal:2018own}. 
In this essay we will use an admittedly simplified setup to 
argue that string-frame de   
Sitter vacua may be realized 
in classical string theory as solutions that are non-perturbative in the (inverse) string tension~$\alpha'$.

Classical string theory can be described by 
two-derivative supergravity theories augmented 
by an infinite number of higher-derivative corrections 
organized 
by the dimensionful parameter~$\alpha'$. 
Such corrections have been determined to a few orders in $\alpha'$ in the 1980s \cite{Gross:1986iv,Gross:1986mw,Metsaev:1987zx}, but a complete determination 
or classification 
in general dimensions is certainly out of reach. While there has been intriguing progress 
in recent years encoding~$\alpha'$   
 corrections in duality covariant  formulations~\cite{Hohm:2013jaa,Hohm:2014xsa,Marques:2015vua,Hohm:2016lge} 
even here a complete description of  higher-derivative
corrections remains out of reach.   
Part of the complication arises because 
in terms of supergravity field variables 
the duality 
transformations themselves acquire derivative corrections making them 
extremely unwieldy.  
Alternatively, when formulated in terms of duality covariant `stringy' field variables, the gauge transformations 
acquire $\alpha'$ corrections, thereby suggesting a novel kind of geometry that, however, 
has only been partially  
 developed.

Had physicists 
classified all  the $\alpha'$ corrections of general $D= d+1$ dimensional theories compatible with duality 
we would then explore the general cosmological solutions by using a suitable time dependent ansatz. 
While 
for now 
this straightforward route is   
blocked,  it turns out that for purely time dependent cosmological solutions
we {\em can} do 
a general analysis.  It is in fact possible to classify all the duality
invariant corrections relevant to these backgrounds.
Here the duality group is $O(d,d,\mathbb{R})$~\cite{Sen:1991zi} and
it efficiently constrains the possible corrections
because, happily, the use of field redefinitions allows us to work with the simple un-corrected duality transformations of the two-derivative theory. 
In this way we circumvent the hard classification problem of the general spacetime theory while getting directly the most general interactions that such a classification
would yield when applied to cosmology. 

In the remainder of this essay we will introduce  the $O(d,d,\mathbb{R})$ 
duality symmetry, state the result of the classification of higher-derivative
cosmological interactions, and use this result
to derive the most general cosmological 
equations for a single scale factor to all orders in $\alpha'$. 
Equipped with these equations we show
how to construct non-perturbative de Sitter solutions.

We start from the $D= d+1$ 
dimensional spacetime action of closed string theory for the universal 
massless fields, 
the metric $g_{\mu\nu}$, the antisymmetric $b$-field $b_{\mu\nu}$, and 
the scalar dilaton~$\phi$, 
\be
\label{lavm}
I \ = \   \int d^Dx  \sqrt{-g}\,  e^{-2\phi} \Bigl( R +  4 (\partial \phi)^2 -\frac{1}{12} H^2 +\frac{1}{4}\alpha' 
\Big(R^{\mu\nu\rho\sigma}R_{\mu\nu\rho\sigma}+\cdots\Big)\Bigr)\;, 
\ee
where $H_{\mu\nu\rho}  =  3  \partial_{[\mu} b_{\nu\rho]}$. Here we have schematically included 
the first $\alpha'$ corrections (with the coefficient for bosonic string theory). 
Let us now 
drop the dependence on all spatial coordinates, leaving 
only the dependence on time $t$, i.e., we set $\partial_i   =   0$, where  $x^\mu   =  (t, x^i)$, $i  =  1,\ldots, d$. 
This is the appropriate 
truncation for cosmology, where 
the non-trivial dynamics 
resides only in the time dependence. 
For the metric, antisymmetric tensor, and dilaton we then set:
\be\label{redansatz}
g_{\mu\nu} \ = \ \begin{pmatrix} -n^2  (t) & 0 \\ 0 & g_{ij} (t) \end{pmatrix}\,, \qquad
b_{\mu\nu} \ = \ \begin{pmatrix} 0 & 0 \\ 0 & b_{ij} (t) \end{pmatrix}\,,   \qquad 
\phi \ = \ \phi (t) \,. 
\ee
Performing this reduction in (\ref{lavm}) to zeroth order in $\alpha'$ one obtains the $O(d,d,\mathbb{R})$ invariant action~\cite{Veneziano:1991ek}:  
\be
\label{vmsrllct} 
\begin{split}
I_0 \ \equiv \ 
  \int d t  \, e^{-\Phi}\,  {1\over n} \,\Bigl(  -\dot \Phi^{\,2} - 
  \frac{1}{8}
  {\rm tr} \big(\dot{\cal S}^2\big) \,  \Bigr) \,, 
\end{split}
\ee
where the `generalized metric'  ${\cal S}$ is a $2d\times 2d$ matrix
constructed  
in terms of $g_{ij}(t)$ and $b_{ij}(t)$: 
\be
\label{etaHdef}
{\cal S} \ \equiv \ 
\begin{pmatrix}  bg^{-1} & g - b g^{-1} b \\[0.7ex]
 g^{-1} & - g^{-1} b 
\end{pmatrix} \;,
\ee
and the $O(d,d)$ invariant dilaton $\Phi$ is constructed in terms
of the scalar dilaton and the metric determinant:    
 \be\label{OddDilaton}
  e^{-\Phi} \ \equiv \ \sqrt{\det g_{ij}} \,e^{-2\phi}\;. 
 \ee
The matrix $\S$ is $O(d,d,\mathbb{R})$ valued and 
satisfies ${\cal S}^2  =  {\bf 1}$. 
The action is manifestly invariant under $O(d,d,\mathbb{R})$ 
transformations:  
\be\label{Sduality}
\S\ \rightarrow \ \S' \ = \  h \,\S \,h^{-1}\;, \qquad  \hbox{with} \qquad  
h\,\eta\, h^t \ = \ \eta\;, \qquad 
\eta \ = \ \begin{pmatrix}  0 & {\bf 1} \\[0.7ex]
 {\bf 1} & 0 
\end{pmatrix}\;.  
\ee
Here $\eta$ is the $O(d,d,\mathbb{R})$ invariant metric. These transformations include, for vanishing $b$-field, 
the duality that sends the metric $g$ to its inverse $g^{-1}$.

Let us next include $\alpha'$ corrections, building on the seminal work of Meissner \cite{Meissner:1996sa} and our own subsequent improvements 
\cite{Hohm:2015doa,Hohm:toAppear}.
In reference~\cite{Hohm:toAppear} we use ${\cal S}^2 = {\bf 1}$ and the most general
duality-covariant redefinitions to classify the higher-derivative corrections
of the two-derivative action~$I_0$.   
Interestingly, no corrections involve the dilaton nontrivially and 
{\em all} corrections can be written in terms of $\dot{\cal S}$, the first order
time derivative of the generalized metric.   
We prove that the general higher-derivative term is constructed
as the trace of even powers of $\dot\S$. 
Multiple traces are allowed but no 
factors ${\tr}(\dot{\S}^2)$ need to be included.  
Thus, the most general duality invariant $\alpha'$ corrected action reads
\be
\label{vmsrllct022} 
\begin{split}
I \ \equiv \    \int d t \,\, e^{-\Phi}  \Bigl(  -\dot \Phi^{2} \ + \  \sum_{k=1}^{\infty}  (\alpha')^{k-1} c_{k}\, \hbox{tr} (\dot{\S}^{2k})
\ + \ \text{multi-traces} \,  \Bigr) \,,  
\end{split}
\ee
where we included the lowest order term in the sum, with $c_1  = -\frac{1}{8}$, 
and picked a gauge with 
lapse function~$n=1$.    
At this stage the $c_k$ with $k\geq 2$ are free coefficients, not determined by duality invariance. The claim is that for any string theory 
there are coefficients $c_k$ (as well as coefficients 
  for the multi-trace terms) so that (\ref{vmsrllct022}) encodes the complete 
$\alpha'$ corrections for cosmological backgrounds. 
Ignoring for simplicity the multi-trace terms,
the equations of motion for $\Phi$, $\S$, and lapse function $n$ are given by   
 \be\label{offshellfunctions}
 \begin{split}
  0 \ &= \ 2\ddot{\Phi} -\dot{\Phi}^2 - \sum_{k=1}^{\infty}(\alpha')^{k-1} c_k \,{\rm tr}\big(\dot{\S}^{2k}\big)\;, \\
  0 \ &= \ \sum_{k=1}^{\infty}(\alpha')^{k-1}k\, c_k \Big(\frac{{d}}{{d}t}\dot{\S}^{2k-1}-\dot{\Phi}\,\dot{\S}^{2k-1}+\S \dot{\S}^{2k}\Big)\;, \\
  0 \ &= \ \dot{\Phi}^2-\sum_{k=1}^{\infty}(\alpha')^{k-1}(2k-1)c_k\,{\rm tr}\big(\dot{\S}^{2k}\big)\;. 
 \end{split}
 \ee
 
We now specialize to the class of Friedmann-Lemaitre-Robertson-Walker (FLRW) backgrounds with curvature $k=0$ and  
vanishing $b$-field. The spacetime metric is then:  
 \be
  ds^2  =  -dt^2  +  a^2(t)\, d{\bf x}^2\;, 
 \ee
where $d{\bf x}^2$ denotes the Euclidean spatial metric 
and $a(t)$ is the scale factor.   
For this metric the generalized metric 
takes the form   
 \be\label{SFRW}
  {\cal S}(t) \ = \ \begin{pmatrix}  0 & a^2(t) \\[0.7ex]
 a^{-2}(t) & 0 
\end{pmatrix}\;.
 \ee

Let us now evaluate 
the equations of motion   
(\ref{offshellfunctions}) for this ansatz in order to determine the $\alpha'$ corrected Friedmann equations. 
It follows from the duality invariance  
(\ref{Sduality}) that these equations are invariant under the `scale factor duality' $a\rightarrow a^{-1}$. 
Let us emphasize that we have in fact obtained  
 the most general equations,  
despite (\ref{offshellfunctions}) being derived from the action without multi-trace contributions. The reason is 
that for the ansatz (\ref{SFRW}) the multi-trace terms give the same structural contributions to the equations of motion
as the single-trace terms, hence resulting only in a `renormalization' of the free parameters $c_k$. 
The generalized Friedmann equations (\ref{offshellfunctions}) 
now become
 \be\label{firstorderEQS} 
  \begin{split}
  \ddot\Phi + \tfrac{1}{2}H f(H) \ &= \ 0\;, \\ 
  \frac{{ d}}{{ d}t}\big(e^{-\Phi} f(H) \big) \ &= \ 0
   \;, \\
   \dot{\Phi}^2+\, g(H) \ &= \ 0\;, 
  \end{split}
 \ee  
where $H(t)$ is the  Hubble parameter:    
 \be
  H(t) \ = \ \frac{\dot{a}(t)}{a(t)}\;, 
 \ee
and $f$ and $g$ are the 
following functions of $H$:  
  \be\label{fDEFFFF}
  \begin{split}
  f(H) \ &\equiv \  d\,  \sum_{k=1}^{\infty}(-\alpha')^{k-1}\, 2^{2(k+1)} \, k\, c_k \,H^{2k-1} \ \ \,\,  = \ -2 d \,  H + \cdots 
  \;, \\
  g(H) \ & \equiv \  d \sum_{k=1}^{\infty}(-\alpha')^{k-1}2^{2k+1}(2k-1) c_k\, H^{2k}  \ = \ - d  \, H^2 + \cdots  \,,
 \end{split} 
 \ee
where we used $c_1 = -\frac{1}{8}$.   
The functions $f(H)$ and $g(H)$ are closely related and  satisfy 
 \be\label{magicidenitty}
  g'(H) \ = \ H f'(H)\;, 
 \ee
a fact that follows from 
one-dimensional reparametrization invariance.  
 One may integrate this equation to express $g$ in terms of $f$, 
  \be\label{gandf}
  g(H) \ = \ Hf(H)-\int_0^H f(H'){d}H'\;, 
 \ee 
for which $g(0)=0$, in agreement with (\ref{fDEFFFF}). 
Solutions of (\ref{firstorderEQS}) perturbative in $\alpha'$ can be found, generalizing the 
early results of Mueller \cite{Mueller:1989in}.

We now ask whether equations~(\ref{firstorderEQS})   
permit de Sitter solutions of the form 
 \be
  ds^2  =  -dt^2  +  e^{2H_0 t} d{\bf x}^2\;, 
 \ee
with constant Hubble parameter $H(t)=H_0$ 
that in absence of matter is related to the cosmological constant by $\Lambda=3H_0^2/c^2$.  
Let us first inspect the equations truncated to zeroth order in $\alpha'$. 
Since then $f(H)\sim H_0$ the second equation of (\ref{firstorderEQS}) implies that $\Phi$ is constant.  
The first equation 
then implies $H_0^2=0$,  
 leading to $H_0=0$ and flat space. Unsurprisingly, the low-energy action (\ref{lavm}) 
to zeroth order in $\alpha'$ thus does not permit de Sitter vacua. 
Nor do the equations truncated to first order in $\alpha'$ allow for de Sitter vacua. 
A non-perturbative de Sitter solution of (\ref{firstorderEQS}) with $H_0\neq 0$ requires, by the second equation, a constant $\Phi$, 
so that with the first and third equation we infer $f(H_0)=g(H_0)=0$. 
A de Sitter solution exists provided~$f$ and~$g$ 
share a common non-vanishing zero. 

To see that these conditions can be met assume that $f$ is given by the sine function, 
 \be\label{fAnsatz}
  f(H) \ \equiv \ -\frac{2d}{\sqrt{\alpha'}}\,\sin\big(\sqrt{\alpha'}\,H\big) \ = \ 
  -2d\, \sum_{k=1}^{\infty}(-\alpha')^{k-1}\frac{1}{(2k-1)!} H^{2k-1}\;. 
 \ee 
From the definition (\ref{fDEFFFF}) it is evident that there are 
coefficients $c_{k\geq 2}$ so that this holds.  
This function has non-vanishing zeroes for $\sqrt{\alpha'}\,H_0 =  n\, \pi$, 
$n\in\mathbb{Z}$.  
The zeroes associated with even, positive $n$ are particularly interesting:   
 \be\label{H_0}
  \sqrt{\alpha'}\,H_0 \ = \ 2\pi\,,\; 4\pi\,,\;\; \ldots\;,  
 \ee
because 
 these are zeroes of $g$ too.  Indeed,  
 with (\ref{gandf}) 
 and taking $\sqrt{\alpha'} H_0=2\pi n $ we have
 \be\label{integrationargument}
  g(H_0)\, = \, -\int_0^{H_0} f(H'){d}H'
  \, = \, {2d\over \sqrt{\alpha'} } \int_0^{2\pi n\over\sqrt{\alpha'}} \sin(\sqrt{\alpha'}H') {d}H' \, = \,  {2d\over \alpha'} \int_0^{2\pi n}\hskip-3pt  \sin u\,  du  \, = \,  0\;. 
 \ee
Equation~(\ref{H_0}) 
 encodes a discrete infinity of de Sitter vacua. 
(We chose $H_0>0$ because the sign is not fixed by   
the equations:  the duality invariance $a\rightarrow a^{-1}$ takes $H\rightarrow -H$.)
Note, however,   
that the natural values of the cosmological constant $\Lambda\sim \frac{1}{\alpha'}$ are many orders of magnitude too large
to account for the observed value, 
which is the cosmological constant problem. 

We do not mean to suggest that (\ref{fAnsatz}) is the function actually appearing in any string theory.  It surely is not.  
But,  
as in (\ref{integrationargument}), one infers that any function $f$ will do the job 
if it has  a non-vanishing zero $H_0$
so that its integral from $0$ to $H_0$ vanishes.  
Alternatively, one can search for an even function $F(H)\sim -H^2$ for $H \to 0$ 
such that it has arguments where $F$ and its derivative $F'$ vanish.  If 
so, $f(H) \equiv F'(H)$ will 
lead to de Sitter vacua.   

Let us emphasize that the above solution is non-perturbative in $\alpha'$. 
There are, however, polynomials $f(H)$ of degree five and higher, 
corresponding to only a finite number of higher-derivative terms, 
that satisfy the above criterion and hence lead to de Sitter vacua.  
This is a familiar  
feature of gravity with a finite number of higher-derivative terms~\cite{Boulware:1985wk}.   
In string theory these vacua cannot be trusted and are almost certainly an artifact of the  truncation.
Reliable de Sitter solutions should either solve the two-derivative equations (and then be corrected perturbatively) or 
be non-perturbative in $\alpha'$, as exhibited above.   
We also stress that the 
above are string-frame de Sitter vacua and are not de Sitter vacua in Einstein frame.

We close with a few remarks on a non-perturbative initial-value formulation of (\ref{firstorderEQS}). 
Suppose we specify  $\Phi(0)$ and $\dot{\Phi}(0)$ at time  $t=0$. 
Then the third equation of (\ref{firstorderEQS}) (the `Hamiltonian constraint') poses constraints 
on the initial value $H_0=H(0)$ of the Hubble parameter: 
 \be\label{Hamiltonian}
  g(H_0) \ = \ -\dot{\Phi}^2(0)\;. 
 \ee
The roots of this equation yield the possible values of $H_0$. Assuming that such a root has been picked, 
(\ref{firstorderEQS}) may then be integrated to later times. Since $g(H)=-d\,H^2+{\cal O}(\alpha')$, the constraint 
(\ref{Hamiltonian}) may always be solved perturbatively, but non-perturbatively there may be many, few 
or no solutions. It has not escaped our notice that this observation has a bearing on the problem of 
how many vacua there are in string theory.

\medskip

 \section*{Acknowledgments} 

We would like to thank Robert Brandenberger,  Maurizio Gasperini, Jean-Luc Lehners, Diego Marques, Krzysztof Meissner, Mark Mueller, 
Ashoke Sen, 
Andrew Tolley and Gabriele Veneziano for useful discussions. 

\noindent
The research of OH is supported by the ERC Consolidator Grant ``Symmetries \& Cosmology".

\end{document}